\documentclass[fleqn,10pt]{wlscirep}

\usepackage{graphicx,psfrag,color}
\usepackage{amssymb,latexsym,mathrsfs,amsmath,amsthm}
\usepackage{subfigure}
\usepackage[utf8]{inputenc}

\title{Disorder and dephasing as control knobs for light transport in optical fiber cavity networks}

\author[1,2]{Silvia Viciani}
\author[2,3,4]{Stefano Gherardini}
\author[1,2]{Manuela Lima}
\author[1,2]{Marco Bellini}
\author[2,3,*]{Filippo Caruso}

\affil[1]{\mbox{CNR-INO, National Institute of Optics,} Largo Fermi 6, I-50125 Firenze, Italy}
\affil[2]{\mbox{LENS, European Laboratory for Non-linear Spectroscopy, University of Florence,} via N. Carrara 1, I-50019 Sesto Fiorentino, Italy}
\affil[3]{\mbox{QSTAR and Department of Physics and Astronomy, University of Florence,} via G. Sansone 1, I-50019 Sesto Fiorentino, Italy}
\affil[4]{\mbox{INFN and Department of Information Engineering, University of Florence,} via S. Marta 3, I-50139 Florence, Italy}
\affil[*]{Corresponding author: filippo.caruso@lens.unifi.it}

\begin{abstract}
Transport phenomena represent a very interdisciplinary topic with applications in many fields of science, such as physics, chemistry, and biology. In this context, the possibility to design a perfectly controllable experimental setup, where to tune and optimize its dynamics parameters, is a challenging but very relevant task to emulate, for instance, the transmission of energy in light harvesting processes. Here, we experimentally build a scalable and controllable transport emulator based on optical fiber cavity networks where the system noise parameters can be finely tuned while maximizing the transfer efficiency. In particular, we demonstrate that disorder and dephasing noise are two control knobs allowing one to play with constructive and destructive interference to optimize the transport paths towards an exit site. These optical setups, on one side, mimic the transport dynamics in natural photosynthetic organisms and, on the other, are very promising platforms to artificially design optimal nanoscale structures for novel, more efficient, clean energy technologies.
\end{abstract}

\begin{document}

\def\be{\begin{equation}}
\def\ee{\end{equation}}
\def\bea{\begin{eqnarray}}
\def\eea{\end{eqnarray}}

\flushbottom
\maketitle
\thispagestyle{empty}

\section*{Introduction}

Very recently several theoretical and experimental studies have shown that the remarkably high efficiency (almost $100\%$) of the excitation energy transfer through a network of chromophores in photosynthetic systems seems to be the result of an intricate interplay between quantum coherence and noise \cite{Lee2007SCI316,Engel2007NAT446,Mohseni2008JCP129, Plenio2008NJP10, Caruso2009JChPh131, Rebentrost2009NJP11,Collini2010NAT463,Chin2010NJP12, Caruso2010PRA81, Caruso2010PRL105,Panitchayangkoon10PNAS107,Hildner2013SCI340,qbiobook}. Indeed, the presence of coherence leads to a very fast delocalization of the excitation that can hence exploit several paths to the target site (named also as sink or reaction center). However, since destructive interference among different pathways and the energy gaps between different sites are obstacles to the transmission of energy, this regime is not optimal by itself. In fact, the additional and unavoidable presence of disorder and noise, which is usually assumed to be deleterious for the transport properties, here instead positively affects the transmission efficiency. This can be explained in terms of the inhibition of destructive interference and the opening of additional pathways for excitation transfer\cite{Caruso2009JChPh131,FC2014,Li2015}. This mechanism, known as Noise-Assisted Transport (NAT), has been recently observed in some physical platforms \cite{Biggerstaff2015, LeonMontiel2015, VicianiPRL2015, FC2016}, but a deep analysis of the underlying contributions, such as interference, disorder and dephasing, is still missing from the experimental point of view.

More in general, the possibility to experimentally realize simple test platforms being able to mimic transport on complex networks and to reproduce NAT effects, would be of high interest from several perspectives. First of all, it would clarify the role of the different disorder, interference and noise contributions in the transport behavior of natural photosynthetic complexes. Secondly, it might represent a model of different complex networks where the role of topology can be further investigated. Finally, one could play with this platform in order to engineer new artificial molecular structures where all these control knobs are optimized to achieve some desired tasks, such as the maximization of the transferred energy or its temporary storage in some part of the network. Such a simple setup would have a remarkable advantage with respect to real biological samples or expensive artificial systems, which are very difficult to manipulate both in their geometry and in the system parameters, since these aspects are often governed by specific bio-chemical laws.

For these reasons, we have recently realized an optical platform, entirely based on fiber-optic components, which has allowed us to provide the first experimental observation of the typical NAT peak in the dependence of the network transmission rate as a function of the amount of noise introduced into the system \cite{VicianiPRL2015}. In such a simple and scalable kind of setup, which has variable and controllable parameters, the coherent propagation of excitons in a $N-$site quantum network is simulated by the propagation of photons in a network of $N$ coupled optical cavities. The optical cavities are based on Fiber Bragg Grating (FBG) resonators \cite{Othons1997RSI68,Hill1997JLT15}, which have been mainly exploited for sensing applications so far \cite{Chow2005JLT23,Gagliardi2010SCI330}, also because they are characterized by a straightforward alignment and easy tunability by tiny deformations of the fiber section within the Bragg mirrors. Each $j$-site of the network (corresponding to each chromophore in a photosynthetic complex) is represented by a FBG resonator, while the local site excitation energy is associated to the resonance frequency of the cavity ($\omega_j$).

Here we provide a complete theoretical and experimental analysis of NAT effects for all the different configurations of our optical model system, demonstrating that constructive and destructive interference, static disorder and dephasing noise can be successfully exploited as feasible control knobs to manipulate the transport behavior of coupled structures and, for example, to optimize the transmission rate.

\section*{Results}
\subsection*{Network of fiber-optic resonators}

\begin{figure}[t]
 \centering
 \includegraphics[width=0.875\textwidth]{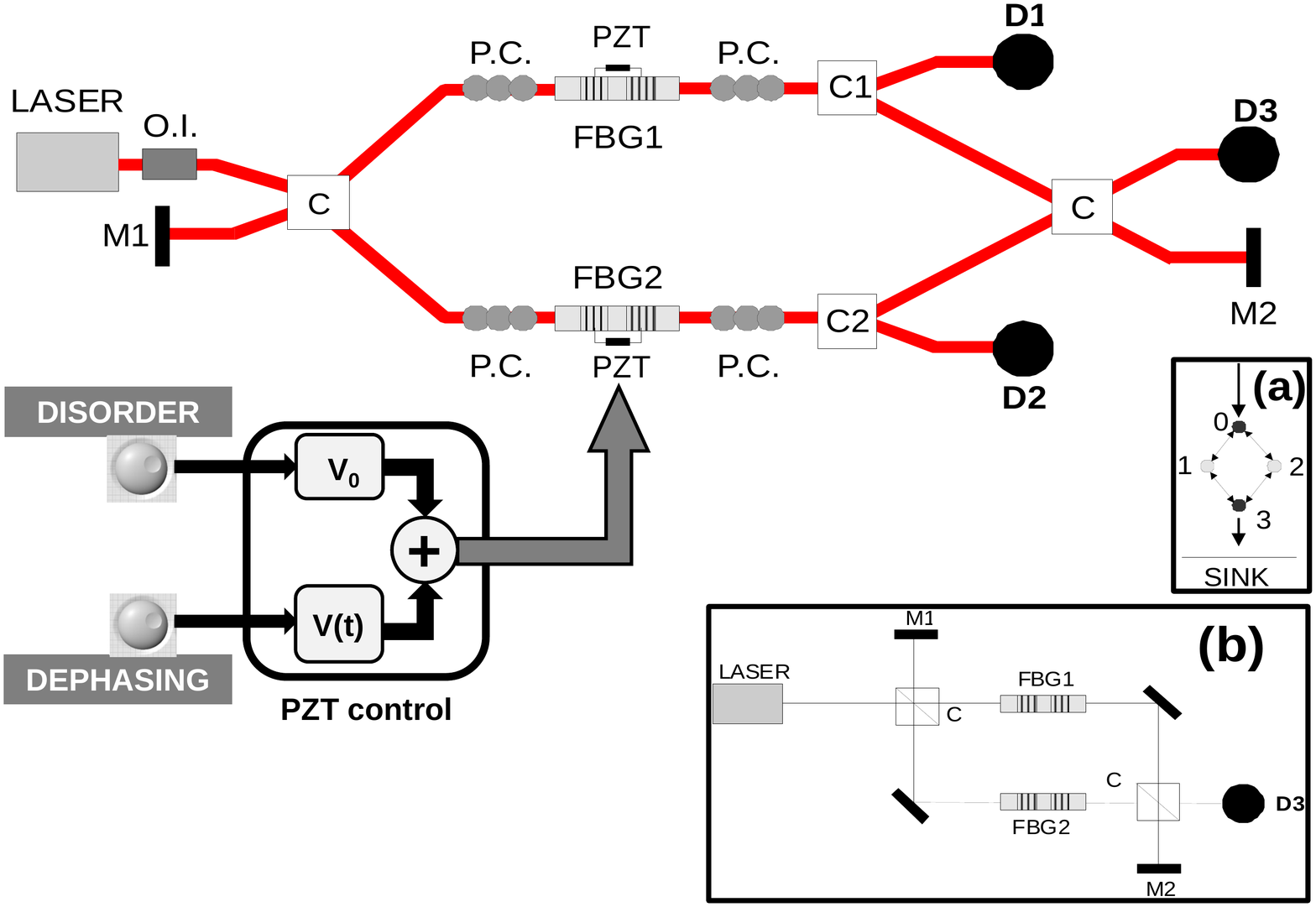}
 \caption{\textbf{Experimental design of the fiber-optic cavity network.} Light is injected in the network by a continuous-wave diode laser, transmitted over coupled optical cavities, and then irreversibly absorbed by a detector measuring the transmission rate. O.I., Optical Isolator; M1 and M2, mirrors; P.C., polarization controller; C, 50x50 fiber coupler; C1 and C2, 90x10 fiber couplers; FBG1 and FBG2, Fiber Bragg Grating resonators; D1, D2 and D3, detectors. Inset (a): scheme of the 4-site network mimicked by the optical setup. Inset (b): simplified scheme of our optical platform.}
 \label{fig1}
\end{figure}

We build up an experimental fiber-optic setup to reproduce energy transport phenomena, including the NAT effect, in the 4-site network shown in the inset (a) of Fig.~\ref{fig1} \cite{VicianiPRL2015}. A detailed scheme of the experimental apparatus is shown in Fig.~\ref{fig1}.
It can be schematically described by the Mach-Zehnder setup of inset (b), which presents two main differences with respect to a standard Mach-Zehnder interferometer: (i) the insertion of a FBG resonator in each path of the interferometer, and (ii) the presence of two additional mirrors (M1 and M2) at the normally unused input and output port of the interferometer. The resonators FBG1 and FBG2 represent the sites 1 and 2 with variable local excitation energy $\omega_1$ and $\omega_2$. The role of the other two sites 0 and 3 is played by two fiber optic couplers (C), which represent two sites with fixed local excitation energy resonant with the energy of the propagating excitation $\omega_\textrm{S}$ ($\omega_0 = \omega_3 = \omega_\textrm{S}$). The presence of the two additional mirrors M1 and M2 makes it possible to couple sites 1 and 2, since the light reflected by the resonators is partially re-inserted into the network by M1, while transmitted light is partially recycled by M2.
The setup is entirely based on single-mode fiber-optic components at telecom wavelength (1550 nm). The choice of fiber components presents several advantages. First, it completely removes issues related to matching the transverse spatial mode of the fields and considerably simplifies the alignment of sources, cavities, and detectors, thus allowing one to easily adjust the network size and topology. Second, working at telecom wavelengths guarantees low optical losses and a low cost of the fiber components. These are two other important factors for the scalability of the apparatus.

The FBG resonators have a finesse of about 100 and a Free Spectral Range of 1~GHz (corresponding to a cavity length of about 10~cm), which result in a full-width at half-maximum linewidth $\textrm{FWHM}_{cav}$ of about 10~MHz. The reflectivity peak of the Bragg gratings is around 1550.13~nm, with a width of 0.7~nm. Each resonator is inserted in a home-made mounting to isolate it from environmental noise and allowing the piece of fiber containing the cavity to be stressed and relaxed in a controlled way by the contact with a piezoelectric transducer (PZT). In such a way the length of each cavity and, consequently, its resonance frequency, can be finely tuned. The input of each resonator is equipped with a fiber polarization controller to allow only one polarization mode to be confined inside the cavity. The light source is a continuous-wave, single-frequency, external-cavity laser (Thorlabs SFL1550S), emitting at 1550~nm,  with a linewidth of about 50~kHz, narrower than the 10~MHz cavity linewidth. The laser source injects light of frequency $\omega_\textrm{S}$ into one input port of the first (50:50) fiber coupler C. Light exiting the two cavities passes through two more polarization controllers before being coupled by a second (50:50) fiber coupler C. Finally, one of the interferometer outputs is measured by detector D3.

All the parameters characterizing the network and that are described in the following section are expressed in terms of the cavity detuning parameter $\Delta x$. This parameter is defined as the difference ($\omega_2 - \omega_1$) between the resonance frequencies of the two cavities in units of their linewidth $\textrm{FWHM}_{cav}$. Two additional fiber couplers (C1 and C2) are used to split a small portion (about 10\%) of the light in each interferometer arm in order to measure the transmission peaks of each single cavity before interference. The transmission signals are measured by detectors D1 and D2 while scanning the laser frequency $\omega_\textrm{S}$ over an interval including a single longitudinal mode of both cavities; from the difference between the frequency positions of the two peaks it is possible to infer $\Delta x$. The three detectors D1, D2 and D3 are biased fiber optic InGaAs photodiodes.

\subsubsection*{Network parameters}

The network configuration can be completely described by 3 parameters: the initial conditions related to global interference, the disorder (static disorder) and the dephasing (dynamical disorder). All these parameter are defined in terms of the previously described cavity detuning parameter $\Delta x$. We theoretically and experimentally characterize the network transmission as a function of these parameters.

\begin{figure}[t]
 \centering
 \includegraphics[width=0.9\textwidth]{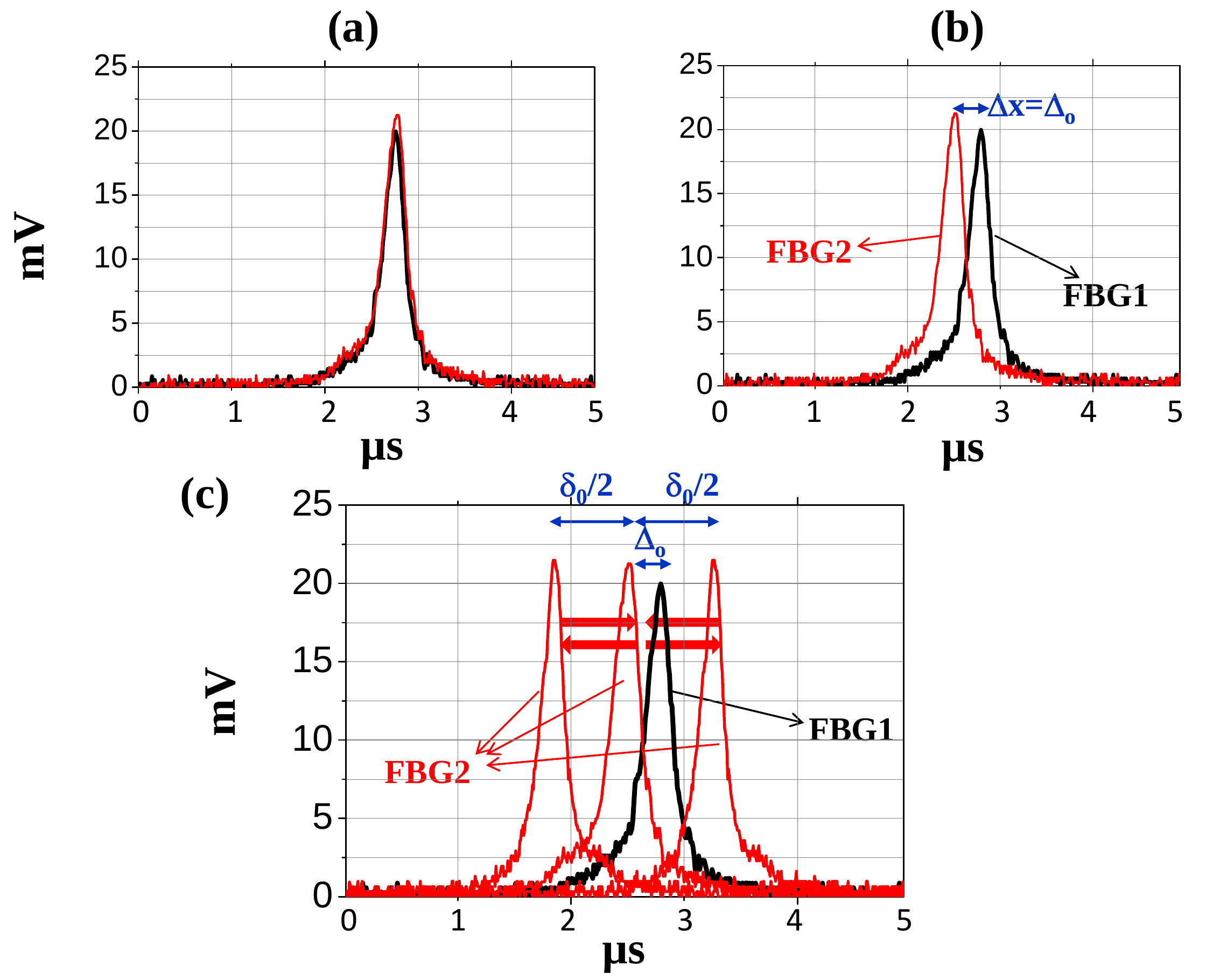}
 \caption{\textbf{Definition of Disorder and Dephasing}.  Transmission peaks of FBG1 (black line) and FBG2 (red line), measured respectively by detector D1 and detector D2  when the laser frequency is scanned over an interval including a single longitudinal mode of both cavities. a) Ordered system without dephasing: $\Delta x = 0$ and constant. b) System with static disorder $\Delta_0$: $\Delta x = \Delta_0$ and constant (no dephasing). c) System with disorder $\Delta_0$ and dephasing $\delta_0$: $\Delta x$ variable between $\Delta_0 - \delta_0 /2$ and  $\Delta_0 + \delta_0 /2$.  }
 \label{fig1bis}
\end{figure}

\textit{Initial Conditions for Interference.-}
Our interferometric apparatus involves no active stabilization. Consequently, the system is intrinsically unstable and the network response will be time dependent (on the time scale of the order of hundreds of ms). This intrinsic instability can be used to establish different initial conditions of global interference for our network. The system throughput when the two cavities are resonant ($\Delta x=0$) and without any kind of noise will vary between a minimum and a maximum value in correspondence of global destructive or constructive interference. The output signal measured by detector D3 in this case will thus set the initial conditions of global interference of the network.

\textit{Disorder or Static Disorder.-}
A network is said to be disordered if the local excitation energies of different sites are unequal ($\omega_j \neq \omega_k$, with $j \neq k $) but constant in time. In our setup, the disorder (or static disorder) of the network can be quantified with the cavity detuning parameter $\Delta x$.
The case of an ordered system is illustrated in Fig.~\ref{fig1bis}a, where $\Delta x = 0$ and constant. The case of a disordered system with a static disorder $\Delta x = \Delta_0 \neq 0$ is shown in Fig.~\ref{fig1bis}b. We assume that a system presents a medium level of disorder if  $\Delta x \sim 1$, i.e. $(\omega_2 - \omega_1) \sim \textrm{FWHM}_{cav}$, and a high level of disorder if  $\Delta x > 1$, i.e. $(\omega_2 - \omega_1) > \textrm{FWHM}_{cav}$.

\textit{Dephasing or Dynamical Disorder.-}
Dephasing (or dynamical disorder) introduces a random phase perturbation in one or more sites of the network, thus resulting in temporal fluctuations of the corresponding resonance frequencies $\omega_j$'s around their stationary values. We can introduce it into our network by slightly changing the value of $\omega_2$ during measurement by means of the PZT. In such a way, $\Delta x$ is not time constant but can vary within an interval of $\pm \delta x / 2$ around $\Delta x$. The amount of dephasing can be quantified by the amplitude $\delta x$ of this interval.
The case of a network with disorder $\Delta_0$ and a dephasing $\delta_0$ is illustrated in Fig.~\ref{fig1bis}c, where $\Delta x$ is variable in the interval $\Delta_0 - \delta_0 / 2$ and $\Delta_0 + \delta_0 / 2$. We assume that a system presents a medium level of dephasing if  $\delta x \sim 1$, and a high level of dephasing if $\delta x > 1$.

\textit{Network Transmission.-}
We define the transmission of the network as the output signal measured by detector D3 in correspondence of the transmission peak of cavity 1 ($\omega = \omega_1$) -- see Methods for technical details on the acquisition procedure. The value of the transmission is normalized to the measured value of the output signal in conditions of constructive interference, without disorder and without dephasing (i.e., for $\Delta x=\delta x=0$).

\subsection*{Theoretical model and numerical results}

To describe the dynamics of our optical cavity network, we consider the following Hamiltonian~\cite{Caruso2011PRA83}:
\be\label{eq1}
\hat{\mathcal{H}}=\sum_{i}\hbar\omega_{i}\hat{a}^{\dagger}_{i}\hat{a}_{i}+\sum_{(i,j)}\hbar g_{ij}\left(\hat{a}^{\dagger}_{i}\hat{a}_{j}
+\hat{a}_{i}\hat{a}^{\dagger}_{j}\right),
\ee
where $\hat{a}_{i}$ and $\hat{a}^{\dagger}_{i}$ are the usual bosonic field operators, that annihilate and create an excitation (a photon in our case) in the $i$-th site of the network, $\omega_{i}$ the corresponding resonance frequency, and $g_{ij}$ are the coupling constants between all the connected sites. The first term in (\ref{eq1}) describes the energy structure of the system, while the second one is related to the hopping process among the network nodes. In the following we will refer to the (random) energy level spacings of the network's sites as \textit{static disorder} and it will be obtained by tuning the frequencies $\omega_{i}$ of the network sites. In addition, the presence of \textit{dephasing} noise, randomizing the photon phase during the dynamics, will be introduced in terms of \textit{dynamical disorder}, i.e. time-dependent random variation of the site energies. We denote with $\gamma_{i}$ the dephasing rate for the site $i$. The latter will be mathematically described by a so-called Lindblad super-operator $\mathcal{L}_{deph}(\hat{\rho})$ , where $\hat{\rho}$ is the density matrix describing the system state that evolves according to the following differential Lindblad (Markovian) master equation:
\be\label{eq2}
\frac{d\hat{\rho}}{dt}=-\frac{i}{\hbar}[\hat{\mathcal{H}},\hat{\rho}] + \mathcal{L}_{deph}(\hat{\rho}) + \mathcal{L}_{inj}(\hat{\rho})
+ \mathcal{L}_{det}(\hat{\rho}),
\ee
where $[\cdot,\cdot]$ denotes the commutator. Moreover, the Lindbladian operators $\mathcal{L}_{inj}(\hat{\rho})$ and $\mathcal{L}_{det}(\hat{\rho})$ in Eq.~\eqref{eq2} describe two distinct irreversible transfer processes, respectively, from the light source to the network (energy injection) and from the exit site to an external sink (energy detection) -- see Methods for technical details. The transferred excitation energy, reaching the sink at time $t$, is defined as
\be\label{eq3}
E_{tr}(t) = 2\Gamma_{det}\int_{0}^{t}\text{Tr}\left[\rho(\tau)\hat{a}^{\dagger}_{k}\hat{a}_{k}\right]d\tau,
\ee
where $\textit{Tr}[\cdot]$ is the trace operation, $\Gamma_{det}$ is the rate at which the photons reach irreversibly the output detector, and $\hat{a}_{k}$ refers to the effective absorption of photons from the site $k$ corresponding to the output site. To compare the theoretical results with the experimental data where the light is continuously injected into the network (with rate $\Gamma_{0}$) and absorbed from the detector, we need to define the \textit{network transmission} as the steady-state rate for the photons in the detector, i.e.
$ \lim_{t \rightarrow \infty} 2\Gamma_{det} \text{Tr}\left[\rho(t)\hat{a}^{\dagger}_{k}\hat{a}_{k}\right]$.
Let us point out that, if one repeats a single-photon experiment many times, one obtains the same statistics corresponding to an injected coherent state \cite{Amselem2009PRL103}, but this holds just because nonlinear processes are not present in our setup. Moreover, note that the model is able to take into account also the slight asymmetry that is present in the experimental setup by imposing different coupling rate in the two paths of the networks, i.e. $g_{01}\neq g_{02}$ and $g_{13}\neq g_{23}$. Such asymmetry is mainly caused by different loss rates in the two resonators.

\begin{figure}[h!]
\centering
\includegraphics[width=.9\textwidth]{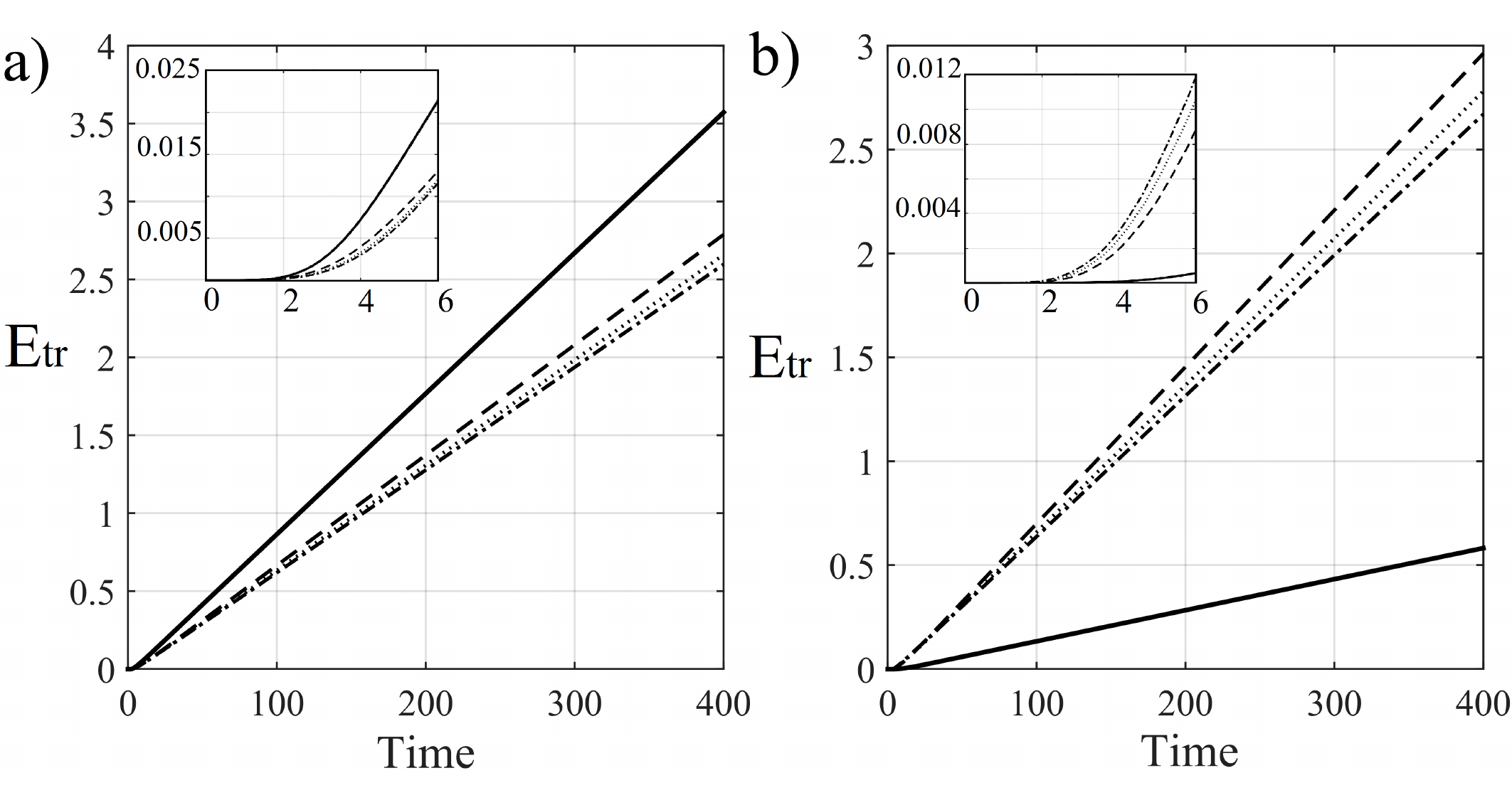}
\caption{\textbf{Time evolution of the excitation transferred energy.} $E_{tr}$ vs time for constructive (a) and destructive (b) interference. In both figures, the curves refers to different values of $\gamma_{2}$ and $\omega_{2}$: $(\gamma_{2},\omega_{2})=[0,0]$ and $(\gamma_{2},\omega_{2})=[0,3]$ for solid and dash-dot lines, respectively, while $(\gamma_{2},\omega_{2})=[2,0]$ and $(\gamma_{2},\omega_{2})=[2,3]$ for dashed and dotted lines. Moreover, we set $\Gamma_{det}=\Gamma_{0}=0.5$. Notice that  $E_{tr}$ is measured in terms of the number of transmitted photons, while the time is in the units of the inverse coupling rates. Inset: Exponential behaviour of the transferred energy in the transient time regime.}
\label{fig:psink2}
\end{figure}
In Fig.~\ref{fig:psink2} different time behaviours of the transferred energy are shown for different initial conditions of global constructive (\ref{fig:psink2}a) and destructive (\ref{fig:psink2}b) interference, and different values of dephasing and static disorder (see Methods for details).
Since the photon injection is continuous in time, the energy in the steady-state condition increases monotonically in time with an asymptotic linear behavior whose slope is indeed the transmission rate. An exponential behaviour is instead observed for the initial temporal regime, as show in the inset of Fig.~\ref{fig:psink2}. Then we find that, for constructive interference both disorder and dephasing individually reduce the transferred energy (no NAT). However, in presence of some disorder inhibiting the path to constructive interference, dephasing slightly assists the transport by opening additional pathways. This can be also intuitively explained by the fact that the two cavities are not energetically on resonance (because of the disorder) but the line broadening effect, induced by dephasing, allows again the hopping between them. On the other side, destructive interference leads to very small transferred energies since the two transmission paths over the two cavities cancel each other (opposite phase). In this case, dephasing and disorder can hinder such a perfect cancellation, thus partially restoring transport, i.e. NAT behaviour.

\subsection*{Experimental results}
The network transmission has been investigated for different initial conditions of global constructive or destructive interference, as a function of both disorder and dephasing on the experimental $4$-site network of fiber-optic resonators in Fig.~\ref{fig1}. In particular, the noise effects on the amount of transferred energy will be shown from two points of view: we analyze the network transmission as a function of dephasing for different values of the disorder, and then as a function of disorder for different values of the dephasing. The experimental data are compared with the theoretical results obtained by the model above in order to show the potential of our setup in emulating transport phenomena in different regimes of interference, disorder and dephasing. The agreement is qualitative because the model is very simplified but still captures the different transport behaviors. Let us remind that in the model the static disorder is added by tuning the cavity frequency of site $2$, $\omega_{2}$, while dephasing is given by $\gamma_{2}$. For both the experimental data and numerical results, the network transmission is normalized to the value of the output signal in the condition of constructive interference without disorder nor dephasing.

\begin{figure*}[h!]
 \centering
 \includegraphics[width=0.72\textwidth]{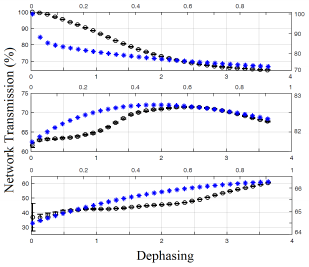}
 \caption{\textbf{Role of dephasing for constructive interference.} (a) Experimental network transmission vs dephasing $\delta x$ for an initial condition of constructive interference and for different values of disorder: no disorder ($\Delta x =0$), medium disorder ($\Delta x =0.7$), and large disorder ($\Delta x =2$); (b) Corresponding numerical evaluation of the  network transmission vs $\gamma_{2}$ when changing disorder $\omega_{2}$.}
 \label{constr_deph}
\end{figure*}

\textit{Constructive Interference.-}
We start by investigating the behavior of the network transmission for initial conditions of global constructive interference.
In Fig.~\ref{constr_deph} the experimental (a) and the theoretical (b) transmission are shown as a function of dephasing for three different configurations of disorder. Without disorder ($\Delta x =0$), the only effect of dephasing is to reduce the transferred energy, i.e. no NAT is observed because the different pathways do already constructively interfere. However, if the system's energy landscape presents some disorder ($\Delta x > 0.4$), NAT effects can be detected and, in particular conditions of disorder, one finds the typical bell-shaped NAT behavior with an optimal value of dephasing that maximizes the network transmission. In other terms, dephasing enhances the transport efficiency if the additional presence of disorder inhibits the otherwise fast constructive-interference path. A similar behavior is found for the theoretical model for parameters compatible with the experimental ones.

\begin{figure*}[h!]
 \centering
 \includegraphics[width=0.7\textwidth]{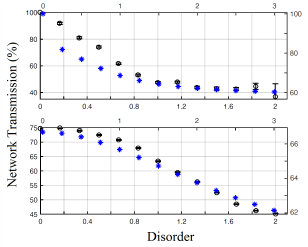}
 \caption{\textbf{Role of disorder for constructive interference.} (a) Experimental network transmission vs disorder $\Delta x$ for an initial condition of constructive interference and for different values of dephasing: no dephasing ($\delta x =0$), and large dephasing ($\delta x = 1.83$); (b) Associated numerical evaluation of the network transmission vs $\omega_{2}$ for different values of $\gamma_{2}$.}
 \label{constr_stat}
\end{figure*}

\begin{figure*}[h!]
 \centering
 \includegraphics[width=0.72\textwidth]{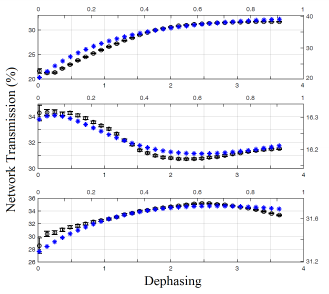}
 \caption{\textbf{Role of dephasing for destructive interference.} (a) Experimental network transmission vs dephasing $\delta x$ for an initial condition of destructive interference and for different values of disorder: no disorder ($\Delta x =0$), medium disorder ($\Delta x =0.7$), and large disorder ($\Delta x =2$); (b) Numerical evaluation of the network transmission vs $\gamma_{2}$ for different $\omega_{2}$.}
 \label{destr_deph}
\end{figure*}

\begin{figure*}[h!]
 \centering
 \includegraphics[width=0.7\textwidth]{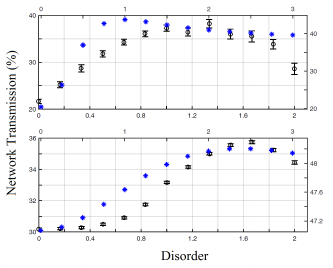}
 \caption{\textbf{Role of disorder for destructive interference.} (a) Experimental network transmission vs disorder $\Delta x$ for an initial condition of destructive interference and for different values of dephasing: no dephasing ($\delta x =0$), and large dephasing ($\delta x = 1.83$); (b) Numerical evaluation of the network transmission vs $\omega_{2}$ when varying $\gamma_{2}$.}
 \label{destr_stat}
\end{figure*}

In Fig.~\ref{constr_stat} the transmission is shown as a function of disorder for two different dephasing configurations. Here, as expected, we find that disorder has a generally negative impact on the transport performance since it always leads to the suppression of the initial constructive interference. However, while a little bit of disorder quickly deteriorates transport in the case without dephasing, the presence of some dephasing noise that broadens the resonances has the effect of making the system more robust against static disorder, with an evidently smoother decay in both the theoretical and experimental cases.
\textit{Destructive Interference.-}
Repeating the analysis above for the case of initial destructive interference, it turns out that both dephasing and disorder independently assist transport, i.e. NAT behavior, since they reduce the amount of interference that prevents the transmission of energy -- see Fig.~\ref{destr_deph} . When the network is in a regime of high-disorder, again the typical bell-like NAT shape is recovered. However, if a lower amount of disorder is included, the additional contribution of dephasing does not further improve the network transmission that instead shows a minimum value for a dephasing value of $\delta x \approx 2.2$.
Finally, in Fig.~\ref{destr_stat} we show the role of disorder in enhancing the transmission rate with its peak moving to higher value of disorder for increasing dephasing values.

\section*{Discussion}
Noise-assisted transport phenomena occur in several physical systems where noise can open additional transport pathways and suppress the ineffective slow ones. In the last years, this scheme has been applied to better understand energy transport in photosynthetic light-harvesting complexes where dephasing noise remarkably enhances the transmission of an electronic excitation from the antenna complex to the reaction center where such energy is further processed \cite{Caruso2009JChPh131}. The basic underlying mechanisms of such behavior are mainly due to line-broadening effects and the suppression of destructive interference, hence the interplay of quantum coherence and noise is responsible for the observed very high transport efficiency. While several theoretical studies have been performed, it is very challenging to actually test these ideas either in the real photosynthetic pigment-protein complex or in artificial ones, since their structure and dynamical properties cannot be controlled or even indirectly measured\cite{Cavities2012,Strong2015} with the required resolution. Moreover, these samples are usually quite expensive or difficult to synthesize. For these reasons, it is very convenient to reproduce such transport phenomena in a controlled system where one can tune the parameters and measure the corresponding dynamical behaviors, while also playing  with the underlying network geometry. This will allow one to better understand the underlying problems and to start engineering new molecular/nano-structures for more efficient and feasible technological applications.

In this work, we have experimentally investigated how to design optimized accessible noise features for energy
transport in a simple and scalable test optical platform, represented by a network of fiber-optic resonators. The
network transmission has been analyzed as a function of static and dynamical disorder in correspondence of different
global interference conditions. In the proposed interferometric apparatus, the effective controllability of parameters is
reached by exploiting the intrinsic instability of the setup, where no phase locking is used for the stabilization of the
system. Therefore, our setup not only allows one to observe the NAT peak in the network transmission as a function of the amount of noise in the system dynamics, but, more importantly, can be exploited to monitor and control the different noise sources to show their individual contribution to the transport dynamics in presence of different types of global interference. Then, we find that, when constructive interference already provides a very fast path to the exit site, dephasing has a detrimental effect and reduces the amount of transferred energy. However, if some disorder is present in the system energy levels, hence blocking the constructive interference path, then dephasing represents a recovery tool to achieve again higher transport efficiency, thanks to NAT effects. On the other side, in presence of destructive interference, both dephasing and disorder are able to speed up the energy transport, with dephasing often providing a faster NAT mechanism. Intuitively, the role of noise in increasing the transferred energy can be explained by considering how the pathways of energy transfer are modified, by destroying the inefficient ones (i.e. inhibition of destructive interferences in the network) or by giving access to more efficient network hubs.

Let us further stress the particular usefulness of the fine controllability of the system parameters and topology in such a simple experimental model compared to the control\cite{Hoyer2014NJP16,FC2012} of the molecular geometries in natural and artificial light-harvesting molecules. This is very hard or even unfeasible in presence of just a few chromophores, with the problem becoming even harder when many chromophores are taken into account. Furthermore, by working with real molecules it is not possible to tune at will their parameters, such as the exciton energies, couplings, noise strength, etc., since the corresponding structure has to be synthesized again each time and just a few molecular configurations are usually available. Our scalable and tunable optical model does instead provide a very useful benchmark platform where all system parameters and geometrical configurations can be very easily changed and individually controlled, even including tens or more of sites.
The power of such optical models is also demonstrated by the fact that they have allowed us to observe, for the first time to the best of our knowledge, the only-theoretically predicted bell-like behavior of the network transmission as a function of the dephasing noise strength \cite{VicianiPRL2015}. Then, increasing the network size, these results are promising in order to observe even more complex transport behaviors that can be very hardly simulated on a computer (see Methods). In particular, optimizing a network for energy transport will correspond to determine the optimal values of the parameters referring to accessible sources of static and dynamical disorder, given the topological features of the network. Notice that in this context the network topology or connectivity plays also a crucial role that can be further investigated by these scalable optical setups.

Therefore, this platform gives us the crucial possibility of testing new complex structures and transport regimes that would not be accessible by studying single molecular systems, but that can be triggered by the observations on these optical models and then synthesized later on artificial molecular devices \cite{Park2016}. Indeed, our theoretical and experimental results show that disorder and dephasing can be exploited as control knobs to manipulate the transport performance of complex networks, e.g. optimizing the final transmission rate or the capability of temporarily storing energy/information in the system. For these reasons, these achievements are expected to find promising bio-inspired applications for practical and much more efficient light-harvesting and energy (information \cite{RMP2014,Caruso2010PRL105}) transport nanotechnologies.

\section*{Methods}
\subsection*{Experimental data acquisition methods}
The network transmission is measured by monitoring the output signal acquired by detector D3 while the laser frequency $\omega_\textrm{S}$ is scanned over an interval comprising a single mode of both cavities. The laser frequency is tuned by applying a voltage ramp, with a frequency of 30~kHz, to the driving current by means of a modulation input port (tuning coefficient about 90~MHz/mA). In order to scan over the cavity modes a variation of less than 1~mA is needed, which yields a residual amplitude modulation smaller than 1\% of a typical transmission peak amplitude. We define the network transmission as the D3 signal for $\omega_\textrm{S}=\omega_1$, where $\omega_1$ can be obtained by the peak of the signal measured by D1. A control software drives the data acquisition chain according to the following steps:\\
(i) $\omega_2$ is initially adjusted by driving the corresponding FBG2 PZT voltage in order to bring the transmission peaks of the two cavities within a few linewidths from each other, thus reducing their detuning $\Delta x$. We monitor the D1 and D2 signals until $\Delta x < 0.1$ (i.e., until the distance of their transmission peaks is smaller than one tenth of their linewidths) and use this condition to trigger the rest of the acquisition; \\
(ii) the value of the D3 signal at this time is recorded and it defines the initial conditions of global constructive or destructive interference for the network transmission;\\
(iii) from this condition, a 1 kHz sinusoidal modulation is applied to the FBG2 PZT so that $\omega_2$ is varied in a range of about $\pm$ 5 linewidths around its initial value. The three signals from D1, D2 and D3 are then acquired continuously for a time of 1 ms so that, for a 30~kHz laser-driving ramp, the signal from D3 allows us to retrieve 30 different values of the network transmission. Since an acquisition time of 1 ms is so short that the entire setup is sufficiently stable to maintain the initial conditions of global interference unchanged, all these acquisitions correspond to the same initial global conditions of interference but to different values of the static disorder parameter $\Delta x$;\\
(iv) after a time acquisition of 1 ms, the acquisition is stopped and the procedure is repeated several times.\\
In such a way, for each initial condition of constructive or destructive interference, a statistically significant dataset of network transmissions (about 14000) has been obtained for different values of $\Delta x$.\\
From this data-set, the transmission for a network with static disorder $\Delta_{0}$ is simply obtained by averaging all the network responses with the same $\Delta x = \Delta_{0}$. An amount  $\delta x$ of dephasing (dynamical disorder) is introduced by averaging the network responses corresponding to different values of $\Delta x$ included in an interval $\pm \delta x / 2$ around $\Delta x$. In particular, the transmission of a network with static disorder $\Delta_{0}$ and dephasing  $\delta_0$, is calculated by averaging all the network transmissions with $\Delta x$ included in the interval $\Delta_0 \pm \delta_0 / 2$.

\subsection*{Theoretical details for the laser source and the detector}
In the model of eq.~\eqref{eq2}, we mimic the experimental conditions by assuming that the network is initially empty, namely with no excitations inside, while a laser source continuously injects photons in the site $0$ with a rate $\Gamma_{0}$. As in Ref. 9, this injection process is modeled by a thermal bath of harmonic oscillators, whose temperature is expressed by the thermal average boson number $n_{th}$ ($n_{th} = 0.1$ in Fig. \ref{fig:psink2}). In the Markov approximation, this process is described by the following Lindbladian term:
\be\label{eq4}
\mathcal{L}_{inj}(\hat{\rho}) = n_{th}\frac{\Gamma_{0}}{2}\left[-\{\hat{a}_{0}\hat{a}^{\dagger}_{0},\hat{\rho}\}+
2\hat{a}^{\dagger}_{0}\hat{\rho}\hat{a}_{0}\right] +(n_{th}+1)\frac{\Gamma_{0}}{2}\left[-\{\hat{a}^{\dagger}_{0}\hat{a}_{0},\hat{\rho}\}+2\hat{a}_{0}\hat{\rho}\hat{a}^{\dagger}_{0}\right],
\ee
where $\hat{a}^{\dagger}_{0}$ represents the bosonic creation operator for the cavity $0$, $\{\cdot,\cdot\}$ denotes the anticommutator, while $\hat{\rho}$ is the physical quantum state of the system.

\begin{figure*}[t]
 \centering
 \includegraphics[width=0.7\textwidth]{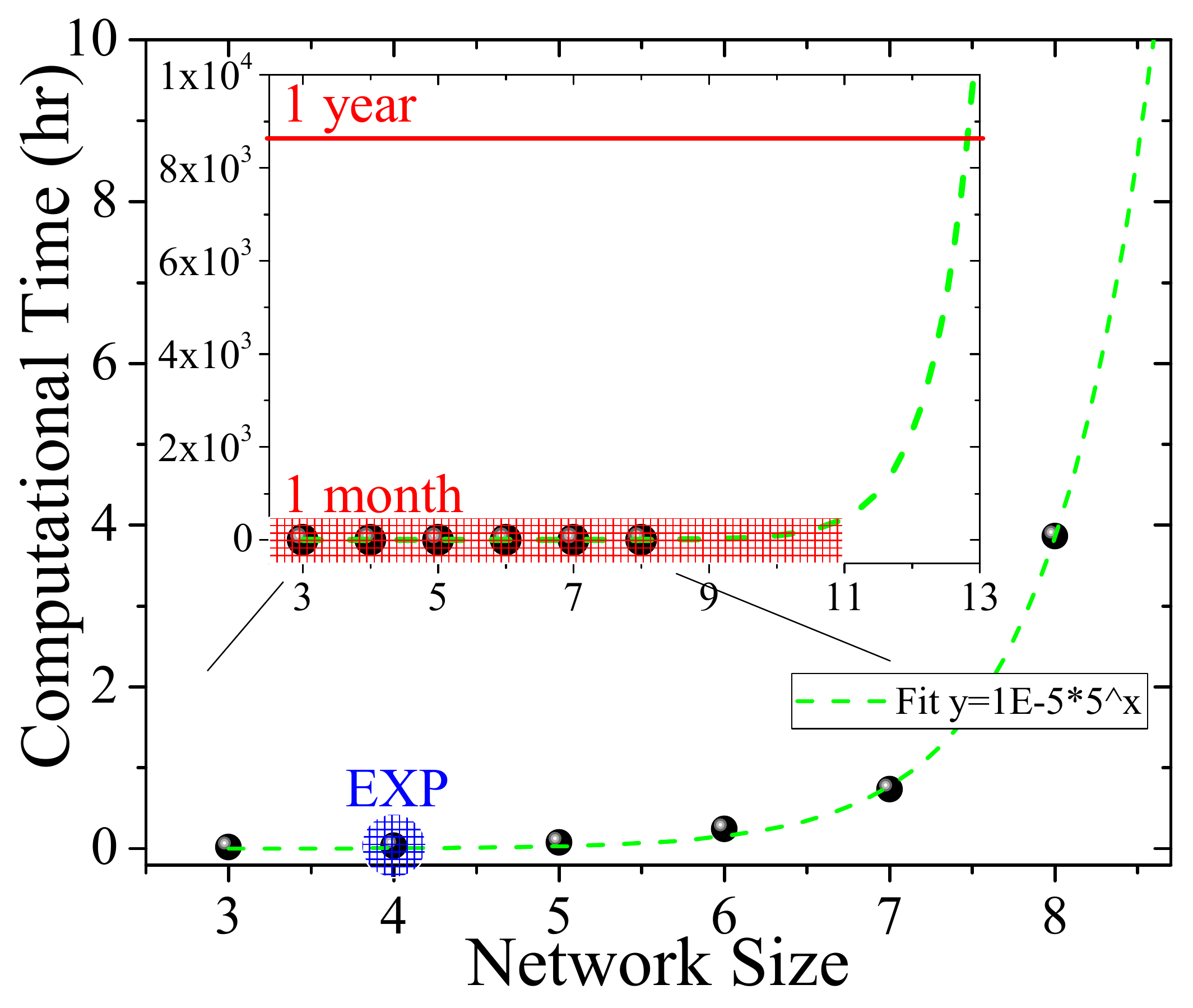}
 \caption{\textbf{Computational complexity increasing the network size.} Computational time (in hours) as a function of the simulated network size (i.e. $N$ number of cavities) for a single realization of the dynamical evolution described in the Main, with only one choice of the system parameters. An exponential behaviour is observed (green dashed line), hence with a scheme of about ten sites/cavities becoming very hard to be simulated by a powerful workstation. Our experimental setup corresponds to the case of four cavities (blue dot) but can be easily extended to more cavities. Inset: Same as above but in a larger scale.}
\label{fig7}
\end{figure*}

The photons leaving the network are detected by the sink, that is usually denoted as the output port of the network, modeling the reaction center of a photosynthetic biological system. Each light-harvesting complex, indeed, is composed by several chromophores that turn photons into excitations and lead them to the reaction center, where the first steps of conversion into a more available form of chemical energy occurs. This part is described by another Lindbladian super-operator, that is
\be\label{eq5}
\mathcal{L}_{det}(\hat{\rho})=\Gamma_{det}\left[2\hat{a}^{\dagger}_{det}\hat{a}_{k}\hat{\rho}\hat{a}^{\dagger}_{k}\hat{a}_{det}-
\{\hat{a}^{\dagger}_{k}\hat{a}_{det}\hat{a}^{\dagger}_{det}\hat{a}_{k},\hat{\rho}\}\right],
\ee
where $\hat{a}^{\dagger}_{det}$ refers to the effective photon creation in the detector with the subsequent absorption of excitations from the site $k$, according to operator $\hat{a}_{k}$, with $\Gamma_{det}$ being the rate at which the photons reach irreversibly the detector.

In order to take into account the experimental imperfections, the non-vanishing coupling constants are set in the range $[0.2, 0.5]$. By varing such parameters, we observe a similar qualitative behaviour, in agreement with the experimental observations, as we simply expect from our very abstract model. The destructive interference is simply obtained by introducing a phase in the hopping strength $g_{01}$, i.e. changing its sign. The cavity resonance frequencies $\omega_{i}$ are all vanishing in the absence of static disorder that instead leads to a variation of the frequency of site $2$ within the range $[0,2]$. Similarly, the only non-zero dephasing rate $\gamma_{2}$ for the cavity $2$ (when dephasing is on) is chosen in the range $[0,1]$, with the dephasing process being in general described by the following Lindbladian super-operator:
\begin{equation}
\mathcal{L}_{deph}(\hat{\rho})=\sum_{i}\gamma_{i}\left[-\{\hat{a}^{\dagger}_{i}\hat{a}_{i},\hat{\rho}\}+
2\hat{a}^{\dagger}_{i}\hat{a}_{i}\hat{\rho}\hat{a}^{\dagger}_{i}\hat{a}_{i}\right] \; .
\end{equation}

\subsection*{Computational complexity}
Here we evaluate the computational time for the single realization of the (numerical) dynamical evolution of networks of increasing size (i.e., number of sites/cavities $N$) with the model as in the Main. The computational complexity does increase exponentially with $N$, as shown in Fig. \ref{fig7}. In other words, already adding a few cavities to our model would take months to theoretically simulate the corresponding dynamics for a given set of parameters, and the computation becomes unfeasible if one wants to reconstruct the figures in our manuscript where thousands of simulations are needed to take into account dynamical disorder, dephasing, etc. In particular, the critical number of sites above which it becomes very hard to reproduce the theoretical data is around $8$ (corresponding to around six months of simulations for one thousand realizations) -- see Fig. \ref{fig7}.
Let us notice that, while the experimental scheme has been realized with coherent states of light that in principle allow its classical simulation time to scale polynomially with the number of optical elements, if one instead considered a full quantum regime (for example, with several single-photon walkers), the computation complexity would have indeed scaled exponentially, as observed above. On the other side, the experimental complexity is not so affected by the network size and, at most, linearly increases in terms of both the cost of the optical components (cavities, beam-splitters, etc.) and the practical realization and observation time of the stationary behavior of the optical system. The use of single-mode telecom optical fiber components is an essential ingredient in this context, as it dramatically reduces costs and losses, besides canceling all issues related to the alignment and spatial mode-matching procedures that should be faced when adding other cavities to the system or changing its topology.

\bibliography{references}

\section*{Acknowledgements}
The authors gratefully acknowledge fruitful discussions with P. Scudo and R. Fusco. This work was supported by the Future in Research (FIRB) Programme of the Italian Ministry of Education, University and Research (MIUR), under the FIRB-MIUR grant agreement No. RBFR10M3SB, and performed in the framework of the ENI contract No. 3500023215. The work of F.C. has been also supported by a Marie Curie Career Integration Grant within the 7th European Community Framework Programme, under the grant agreement QuantumBioTech No. 293449.

\section*{Author Contributions}

F.C., S.V. and M.B. conceived and supervised the whole project; S.V. and M.L. planned and carried out the experiment with input from M.B. and F.C.; F.C. led the theory; S.G. and F.C. performed the numerical simulations. All authors contributed to the discussion, analysis of the results and the writing and reviewing of the manuscript.

\section*{Additional Information}
The authors declare no competing financial interests.

\end{document}